# Refractive index mediated plasmon hybridization in an array of aluminium nanoparticles

Alina Muravitskaya,*[a,b] Anisha Gokarna,[a] Artur Movsesyan,[a] Sergei Kostcheev,[a] Anna Rumyantseva,[a] Christophe Couteau,[a] Gilles Lerondel,[a] Anne-Laure Baudrion,[a] Sergey Gaponenko[b] and Pierre-Michel Adam[a]

The arrangement of plasmonic nanoparticles in a non-symmetrical environment can feature the far-field and/or near-field interactions depending on the distance between the objects. In this work, we study the hybridization of three intrinsic plasmonic modes (dipolar, quadrupolar and hexapolar modes) sustained by one elliptical aluminium nanocylinder, as well as behavior of the hybridized modes when the nanoparticles are organized in array or when the refractive index of the surrounding medium is changed. The position and the intensity of these hybridized modes were shown to be affected by the near-field and far-field interactions between the nanoparticles. In this work, two hybridized modes were tuned in the UV spectral range to spectrally coincide with the intrinsic interband excitation and emission bands of ZnO nanocrystals. The refractive index of the ZnO nanocrystals layer influences the positions of the plasmonic modes and increases the role of the superstrate medium, which in turn results in the appearance of two separate modes in the small spectral region. Hence, the enhancement of ZnO nanocrystals photoluminescence benefits from the simultaneous excitation and emission enhancements.

## Introduction

The use of aluminium in plasmonics and in particular for the fluorescence enhancement and surface enhanced Raman scattering (SERS) has received significant attention in recent years. In the UV range, enhancement of the secondary light emission by the well-known plasmonic metals, Au and Ag, is not efficient because of the high absorption.[1–3] Unlike in these materials, the interband transitions of aluminium are located in the near-infrared.[4,5] Also, aluminium is a widespread abundant material and has a favourable dielectric permittivity to sustain the localized surface plasmon (LSP) resonances in the deep-UV-vis range.[4–7] By varying the shape and size of Al nanoparticles one can tune the plasmonic resonance in a wide range and even excite multiple modes in different spectral regions simultaneously.[8] Despite the fact that a few nm-thick native oxide layer grows at the surface in the ambient condition and affects the surface enhancement processes, there is still a number of articles illustrating the use of the Al substrates and nanoparticles for SERS and fluorescence enhancement.[1,9–17]

Optical properties of plasmonic nanoparticles arrays differ significantly from the properties of isolated nanoparticles, due to the appearance of lattice resonances (LR) resulting from the diffraction mediation of the radiative coupling between nanoobjects.[18] The position of the diffraction lattice modes depends on the array spacing, refractive index of the surrounding medium and the illumination angle.[6,18–23] These collective resonances are usually considered to originate from the interaction between the individual dipolar modes of each nanoparticle. Lattice resonances originating from the quadrupolar or gap modes were also demonstrated.[21,24,25]

Besides the far-field coupling due to the diffraction in the plane of the array, complex plasmonic nanoparticles or nanoparticle clusters can benefit from the near-field coupling or hybridization of the intrinsic modes,[26–28] which can result in changes of the LSP position and the appearance of Fano resonances. Similar behaviour is observed in non-symmetrical environments.[26,29–31] For example, one can create conditions for the effective coupling by introducing symmetry breaking into the system by means of a dielectric or a metal substrate.[28,32,33] However, the simultaneous control of the far-field and near-field coupling in one system still requires further investigation.

Low-dimensional ZnO-based materials continue to attract interest due to their highly sought-after properties suitable for optoelectronics such as a wide bandgap of 3.37 eV, large excitonic binding energy of 60 meV and size dependent optical and electronic properties.[34,35] Recently the improvement of the near band gap emission of ZnO nanostructures with the help of different plasmonic materials has become a developing research field. At first it was achieved using silver and gold[36–39] and then the focus was shifted towards aluminium, since it is more commercially viable material.[40–45]

In this work, we study the plasmonic modes hybridization in

[a.] Laboratory Light, nanomaterials & nanotechnologies (L2n), CNRS ERL 7004, University of Technology of Troyes, 12 rue Marie Curie, 10004 Troyes Cedex, France. E-mail: alina.muravitskaya@gmail.com
[b.] B. I. Stepanov Institute of Physics, National Academy of Sciences of Belarus, Nezavisimosti 68-2, Minsk 220072 Belarus.
† See DOI:

an array of elliptical Al nanocylinders. Due to its elliptical shape, a nanocylinder has a number of excited plasmonic modes such as dipolar, quadrupolar and hexapolar modes, which interact with each other in an asymmetric environment. The energy of the hybridized modes was shown to depend on the far-field interaction of the nanoparticles in array. Under certain conditions, two plasmonic modes appear in the UV spectral range and spectrally coincide with the intrinsic interband excitation and emission bands of ZnO nanocrystals.

## Experiment and methods

The ordered arrays of elliptical aluminium nanocylinders were fabricated by electron beam lithography (Raith eLINE system), which enables the production of nanostructures having well-controlled size, shape, and interparticle spacing. The thickness of aluminium was set to 50 nm and interparticle spacing was 216 nm (Fig. 1a). Structural characterization of these samples was conducted using a scanning electron microscope (SEM) (Fig. 1d).

ZnO nanocrystals precursor was prepared by dissolving zinc acetate in ethanol (0.48M) under stirring for 24 h. The solution was then uniformly deposited on a glass substrate by spin coating followed by the annealing on a hot plate for 5 min at 300°C. The annealing leads to transformation of amorphous ZnO to polycrystalline phase with the crystals of a mean diameter of 5-7 nm. Reference sample (identic substrate with bare Al nanocylinders) was prepared and heated the same way up to 300°C. The plasmonic peaks did not shift, only a variation of the intensity was observed.

Optical characterization of the samples was performed at room temperature. For the extinction measurements the transmitted light was collected using a collection lens coupled to an optical fiber and a spectrometer Ocean Optics USB2000+. Photoluminescence spectra were recorded by a spectrometer under the excitation by the He-Cd laser (325 nm).

Numerical simulations were performed using Lumerical software, based on the Finite Difference Time Domain method. ZnO was simulated as a layer with a known refractive index (about 1.7 for the studied spectral range) and extinction coefficient.[46] The surface charges distribution was calculated according to the formula described elsewhere.[47]

## Results and discussion

LSP resonance frequency of the fabricated aluminium nanostructures is mainly determined by the size and shape of the nanoparticles, and also by the dielectric permittivity of the substrate and the ambient medium. The coupling between the lattice modes of the pattern and the plasmonic modes can modify the shape and position of the resonance peak.[18] Al nanoparticles sustain LSP resonances in a wide spectral range. For single cylindrical nanoparticles, the main in-plane dipolar resonance shifts from 300 to 600 nm as their diameter increases from 70 to 180 nm.[4] One of the main requirements for the effective coupling between LR and LSP modes is their spectral overlap.

We studied elliptical aluminium nanocylinders (major and minor axes experimentally measured of 192 nm and 102 nm, respectively) in an array having a pitch of 216 nm. Depending on the light polarization the elliptical nanocylinders can sustain dipolar resonances of different energies along the two axes. For the excitation polarized along the minor axis (TE) of the nanocylinders the extinction spectrum has a maximum at 360 nm (Fig. 1b). One can notice that this peak is slightly asymmetrical. Indeed, for a pitch of 216 nm the position of the (±1, 0) Rayleigh is around 334 nm. Then, the interaction between the LSP resonance of nanocylinders and the Rayleigh anomaly results in the narrowing of the resonance and the asymmetry of the curve. The extinction spectrum for the polarization along the long axis (TM) features a wide plasmonic maximum near 490 nm. As the LSP resonance and the LR are spectrally separated, the narrowing observed for the TE polarization does not appear here.

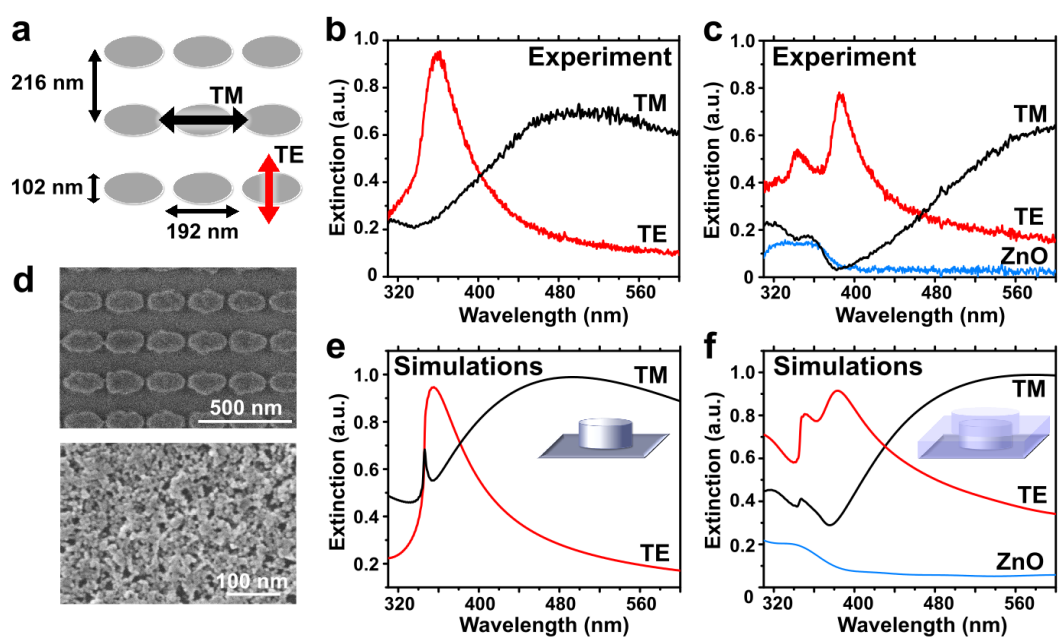

**Fig. 1** Schematic illustrations of the array geometry (a). Extinction spectra of the Al array (two polarizations) without (b) and with (c) overlying layer of ZnO nanocrystals for the polarizations along the short (TE) or long (TM) axes of the elliptical cylinders and the reference spectrum of the ZnO nanocrystals. SEM images of the elliptical nanocylinders array and ZnO nanocrystals layer (d). Corresponding numerical simulations for extinction measurements (e, f).

The results of the FDTD numerical simulations for the proposed system are shown in Fig. 1e. The nanoparticles were simulated as an array of aluminium elliptical nanocylinders surrounded by a 3 nm-thick oxide layer and deposited on a glass substrate. Taking into account that the simulations are performed for an infinite array, the calculated extinction spectrum agrees well with the experimental results. Indeed, the optical response of a finite and infinite arrays may differ.[23] The calculated spectrum of the array for the TM polarization has a wide maximum around 480 nm, but it also has a small narrow band at 350 nm which matches the LR position. This peak does not appear in the experimental spectrum probably due to the imperfections of nanocylinders shape and the finite size of the experimental array.

The extinction spectra of the Al array covered with ZnO nanocrystals is shown in Fig. 1c. In the case of TM polarization (black curve), the peak maximum is red-shifted by 100 nm, and less intense extinction bands appear in the range of 300-370 nm. To avoid the contribution of the absorption band of the ZnO nanocrystals, all spectra obtained for the coated arrays were normalized by the extinction spectrum of the ZnO nanocrystals without Al nanoparticles (Fig. 1c blue curve).

Two distinct peaks at 342 and 386 nm appear for the polarization along the short axis of the nanocylinders (Fig. 1c). The splitting of the extinction peak after the emitter deposition is often associated with strong plasmon-exciton coupling, when the surface plasmon mode coherently hybridizes with emitter excitons.[48,49] Also recently, the coupling of excitons with lattice plasmons was studied in weak coupling regime.[50] However, two maxima at 335 nm and 370 nm appear for the numerical simulation as well (Fig. 1f), although in our simulations we do not consider ZnO as an emitter but only as a layer with known refractive index. Then, different factors affecting the spectra were considered to understand the origin and behavior of the new modes.

**Single nanocylinder modes.**

We perform the numerical study of a single elliptical Al nanocylinder with a long axis of 186 nm and a short axis of 96 nm (same sizes as mentioned previously but without native oxide layer) in air (Fig. 2). One can see that single Al nanocylinder extinction spectrum has two distinct bands peaking at 210 and 310 nm respectively, and a shoulder at 200 nm (Fig. 2a). The peak at 310 nm is wide and has dipolar features, while the high frequency mode is narrower. The absorption spectrum features two distinct peaks (195 nm and 215 nm), which merge into a single one on the extinction spectrum. These two maxima are associated with the out-of-plane quadrupolar mode excited due to the phase retardation

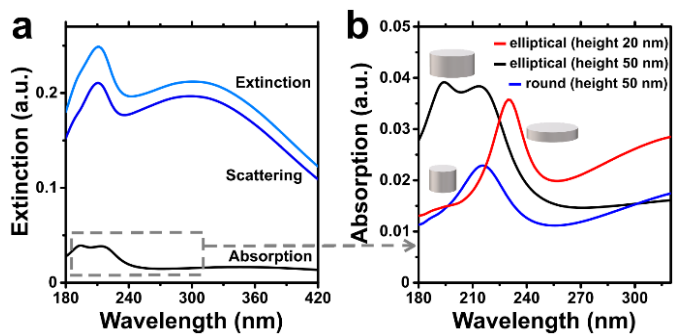

**Fig. 2** (a) Calculated absorption, scattering and extinction spectra of the Al elliptical nanocylinder in air (polarization along the short axis). (b) Sections of absorption spectra of elliptical and circular Al nanocylinders.

effect, and in-plane hexapolar mode, excited due to the elliptical shape and size of the nanocylinder (black curve in Fig. 2b). This assumption is confirmed by the separate simulation for an elliptical nanocylinder with a height of 20 nm which is not high enough to be influenced by the retardation effect under normal incidence excitation (red curve in Fig. 2b). The spectrum exhibits one peak (below 300 nm) at 230 nm that can be attributed to the hexapolar mode. Simulation performed for the circular cylinder (blue curve in Fig. 2b) shows maximum of absorption at 215 nm arising from the quadrupolar mode excitation due to the retardation effect. It should be noted that the position of the hexapolar peak of elliptic nanocylinder is blue-shifted for the 50 nm-high nanocylinder due to the known effect when the LSPR wavelengths strongly blue-shift upon the increase of the nanoparticle height (and thus decrease of the aspect ratio).[51] As it was pointed earlier, Al nanoparticles oxidize in air and the 3 nm thick $Al_2O_3$ layer appears on the surface[4] and induces a red shift of the spectra since the average refractive index around a metal nanoparticle increases (spectrum not shown). The extinction maxima in this case are shifted to 225 nm and 320 nm, respectively.

**Substrate influence and near-field interactions.**

In all our experiments, the elliptical nanocylinders were organized in a dense array on a substrate. The substrate breaks the symmetry of the system and induces a significant shift of both plasmonic modes. The resulting extinction peaks arise at 255 nm and at 390 nm, respectively (Fig. 3a). Similar peaks were found to appear in the spectra of a circular cylinder of 102 nm in diameter (Fig. 3b). In this case there are two hybridized modes which are dipolar/quadrupolar in nature according to the charge distribution calculations (Fig. 3d). These modes were observed as well in silver nanocubes and nanocylinders[29,33] and were found to be induced by the symmetry breaking of the surrounding medium. The mode at 400 nm is mostly localized near the glass surface and can be described as a sum mode ($D_0+Q_0$), where $D_0$ is a dipolar mode of the nanocylinder in air and $Q_0$ is the quadrupolar mode. For the mode at 260 nm, the electric field is confined on the top plane of the nanocylinder and can be explained as a difference of the initial modes ($D_0-Q_0$). The charge distributions are different for the elliptical (Fig. 3c) and circular nanocylinders (Fig. 3d). For the latter, the simulations show dipolar-like charge distribution for both modes, while for the elliptical nanocylinders the modes are similar to the mixture of hexapolar mode and already hybridized dipolar/quadrupolar modes.

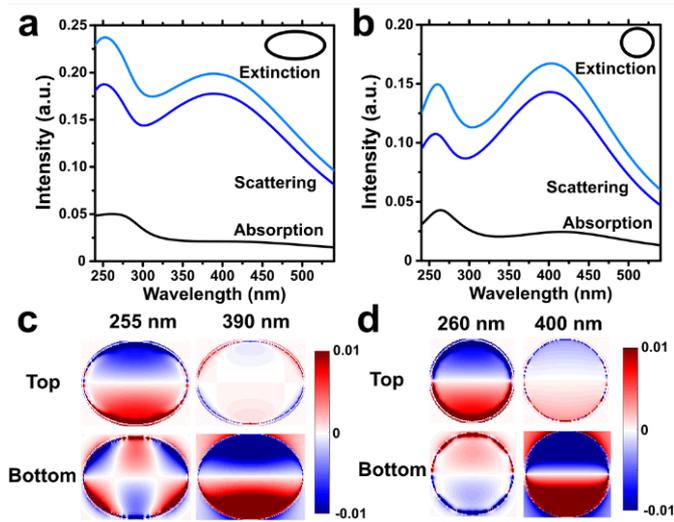

**Fig. 3** Absorption, scattering and extinction spectra of a single Al nanocylinders with 3 nm of $Al_2O_3$ deposited on a substrate (polarization along the short axis (TE)) of an elliptical (a) and (b) circular shape. The charge distributions for the resonance wavelengths for elliptical (c) and circular (d) nanocylinders.

In the array of the elliptical nanocylinders, the length of the major axis is very close to the interparticle distance, which results in a small (24 nm) gap between the nanocylinders along their long axis and in the enhancement of the local electric fields. It was shown that at this distance, near-field interactions influence the spectrum of the structure.[52] Figure 4a shows the spectra of a single elliptical nanocylinder and five nanocylinders deposited on the substrate with 24 nm interparticle distance. Due to the interparticle interactions the main band experiences a blue shift to 354 nm.[53] We also conducted additional studies wherein a layer of ZnO was introduced into the system. The ZnO deposition induces a redshift of the whole spectrum (Fig. 4b) due to the local refractive index change. The peaks become closer and the main maxima are located at 305 nm and 410 nm, respectively.

**Origin of the modes.**

Simulations results for the infinite array of Al nanocylinders covered by layer of ZnO nanocrystals (Fig. 1(f)) show two

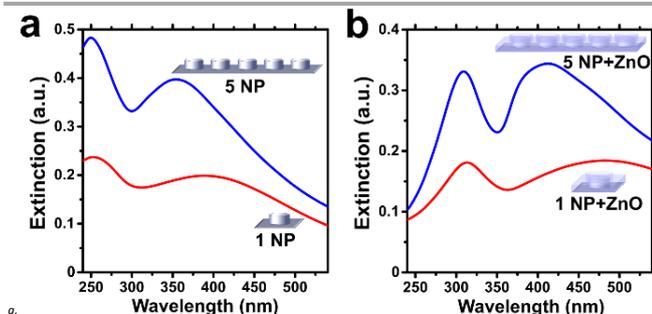

**Fig. 4** Extinction spectra of a single Al nanocylinder and 5 Al nanocylinders separated by 24 nm (polarization along the short axis) deposited on a glass substrate in air (a) and with overlying layer of ZnO nanocrystals (b).

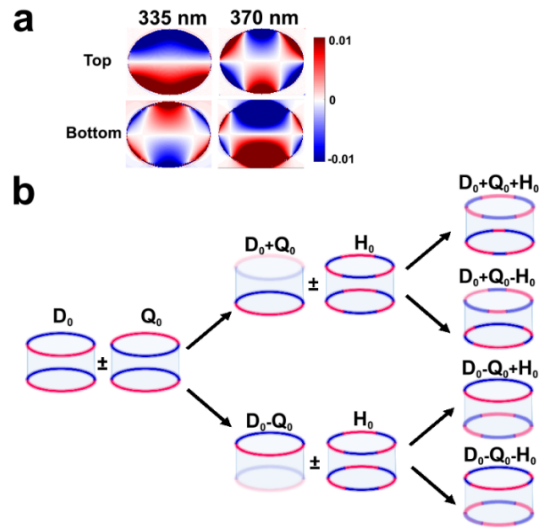

**Fig. 5** (a) Results of the numerical simulations for an infinite array of Al elliptical nanocylinder showing the charge distributions at 335 nm and 370 nm, respectively, for one nanocylinder and (b) schematic of the hybridization of the modes.

modes: at 335 nm and at 370 nm. For both of them one can notice the different charge distributions at the top and at the bottom of the nanocylinders (Fig. 5a). The mode at 335 nm is similar to the mode of the single elliptical nanocylinder at 255 nm (Fig. 3c). In vertical cross section it has the features of the quadrupolar mode, while the top plane distribution is similar to the dipolar mode and the bottom plane distribution is similar to the hexapolar mode. The mode at 370 nm resembles a non-symmetrical hexapolar mode. It has some similarities with the mode for the single elliptical nanocylinder on the substrate at 390 nm (Fig. 3c), although hexapolar component appears to be more intensive, due to the nanocylinders packing and higher refractive index of the superstrate with ZnO deposition. The appearing of these two modes can be explained with the help of a hybridization model. The initial modes in the system are dipolar, quadrupolar and hexapolar ones (Fig. 5b). In a nonsymmetrical environment, these modes interact with each other. At first, we consider the coupling of the dipolar and quadrupolar modes that results in formation of two new modes, namely: $D_0-Q_0$ and $D_0+Q_0$. Each of these modes interact with the hexapolar mode. Further on, the mode at 370 nm is similar to the $D_0+Q_0-H_0$ mode and the mode at 335 nm is similar to the $D_0-Q_0+H_0$ mode, where the $D_0$, $Q_0$, $H_0$ are the dipolar, quadrupolar and hexapolar modes of the nanocylinder in air, respectively.

**Superstrate and pitch influence.**

The superstrate media are known to change the LR positions as well as affect the intrinsic LSP resonance position which is widely used in LSP resonance based refractive index sensors.[6,54–58] The typical behavior of the plasmonic modes is a red shift upon the increase of the surrounding refractive index. We numerically studied the circular nanocylinders with a diameter of 102 nm and 50 nm height arranged in an array with a 216 nm pitch on the glass substrate (Fig. 6a). Herein, the refractive index of the superstrate is varied from 1 to 2. The main peak shifts from 360 nm to 500 nm when the superstrate refractive index rises. The shift and broadening of the peak result from the decoupling of the LSP and lattice resonance. The LR excited in the substrate (at 335 nm) becomes less pronounced and tends to vanish with the rise of the superstrate refractive index.

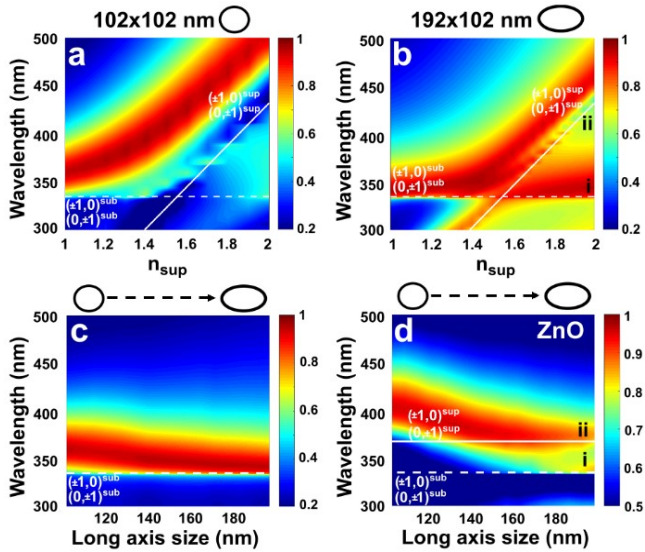

**Fig. 6** Extinction spectral maps versus a superstrate refractive index for a circular nanocylinder $d$=102 nm (a) and an elliptical nanocylinder (192x102 nm) (b). The extinction maps depending on the length of the nanocylinders long axis without (c) and with (d) overlying layer of ZnO nanocrystals.

Interesting features appear in the spectra of the elliptical nanocylinders (Fig. 6b) with the same short axis (102 nm), but different long axis (192 nm). The incident light is polarized along the short axis of the nanocylinders. The extinction spectrum changes significantly with an increase in the superstrate refractive index (Fig. 6b) and shows new features, when compared to the case of the circular nanocylinder (Fig. 6a). For $n_{sup}$<1.5 the mode at 340 nm remains almost unchanged but the additional mode of the higher energy shifts strongly with the rise of the refractive index of the superstrate. For $n_{sup}$>1.5 mode (ii) shifts toward the red region, while mode (i) stays coupled with the LR of the substrate. Mode (ii) corresponds to previously described mode $D_0-Q_0+H_0$, and mode (i) is associated with mode $D_0+Q_0-H_0$.

The effect of the elongation of a nanocylinder axis on the spectral behavior of the array was also analyzed (Fig. 6c). When the superstrate is air and the incident wave is polarized along the short axis, the spectrum does not change significantly with elongation. The main resonance becomes more narrow as the length of the long axis increases. The extinction behavior of the array covered with ZnO nanocrystals is different (Fig.6d). The refractive index of ZnO nanocrystals layer depends on the wavelength and is about 1.7 for this spectral range. Extinction spectra for the elliptical nanocylinder array features two maxima at 335 nm and 370 nm. Both of them coincide with the diffraction order wavelengths in the substrate and the superstrate. The origin of the peak at 335 nm becomes clear from Fig. 6b. This mode shifts from the deep-UV region and couples with the LR of the substrate. In turn the mode (ii) shifts to the red and interacts with the lattice resonance of the superstrate. The two modes are observed only when the long axis is longer than 170 nm and hexapolar mode influence becomes more pronounced.

A change in the periodicity also influences the mode appearing (Fig. 7). For the air superstrate, the resonances are shifted to the red following the LR of the substrate (Fig. 7a). The ZnO superstrate induces the two modes which appear for the pitch of 200-240 nm in the spectral range between 300 nm and 400 nm. Fig. 6b shows that in the case of a larger pitch, the LSP mode from the deep-UV does not overlap with a diffraction order frequency and therefore does not couple with it. However, one can assume that if the superstrate refractive index increases, an additional mode will be observed for a larger range of periods. These results show that it is possible to observe two lattice modes of the first order in the substrate and the superstrate if these grating-induced resonances are supported by the nanoparticles plasmonic modes.

**Photoluminescence enhancement.**

Al nanostructures were found to influence ZnO photoluminescence (PL). Upon the excitation at 325 nm the PL spectrum features a peak at 385 nm which corresponds to a narrow near band-edge emission of ZnO. The experimental measurements show a 9.7-fold enhancement of the ZnO PL on the described array of Al nanostructures compared to bare ZnO nanocrystals (Fig. 8). The plasmonic enhancement of the photoluminescence is associated with a strong electric field localization as well as change in the radiative and non-radiative decay rates of the emitter in the vicinity of the plasmonic nanoparticle.[58–61] In our configuration the native aluminium oxide layer plays a role of a spacer between the ZnO nanocrystals and aluminium. Previously the emission of ZnO near the metal nanoparticles was shown to increase significantly with the introduction of a spacer into the system due to the decrease in the nonradiative relaxations of the excited state.[40] Aluminium nanoparticles do not change the photoluminescence spectra shape, which can serve as an evidence of the absence of the strong coupling in the system.

We study the effect of the plasmonic mode appearing in the extinction spectrum of Al nanocylinders at 342 nm on the enhancement of ZnO nanocrystals photoluminescence. We experimentally considered the array having close characteristics to previous nanocylinders (sizes 170 nm x 102 nm). The array period was fixed at 216 nm. The extinction spectra of these arrays in air is almost the same. For Al nanocylinders covered with a ZnO nanocrystals, two distinct peaks at 342 nm and 386 nm appear only for the array with bigger nanoparticles (inset in Fig. 8), which is in agreement with the simulation (Fig. 6d). For the nanocylinders 170 nm x 102 nm, the peak at 342 nm does not appear clearly.

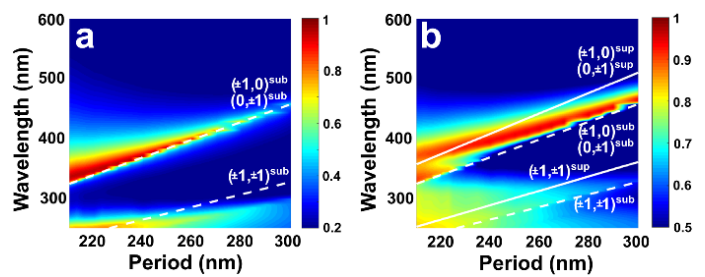

**Fig. 7** Dependence of the extinction on the period of the array of elliptical Al nanocylinders for air superstrate refractive index (a) and ZnO nanocrystals layer (b). Diffractive orders are marked with white lines.

PL of ZnO nanocrystals on the 170 nm x 102 nm nanocylinders was 5.4-fold enhanced in comparison to their PL on the glass substrate. It is important to note that the area of Al is 15 % lower in the case of smaller nanocylinders. The increase in area alone cannot account for the rise in the enhancement factor from the 5.4 times to 9.7. We can conclude that the PL signal of ZnO nanocrystals benefits from the additional enhancement of the excitation when the mode at 342 nm appears.

## Conclusions

In summary, we have performed an experimental and numerical study of the plasmonic modes hybridization in an array of the elliptically shaped aluminium nanocylinders. Aluminium nanoparticles sustain the localized surface plasmon resonances in the wide spectral region that allows in detail analysis of the variety of the excited modes and their interaction in the UV region. We found that the appearance of the two plasmonic maxima in the extinction spectrum after the deposition of the emitter, which is usually associated with a process of a strong plasmon-exciton coupling, may originate from the shift and interaction of the modes with lattice resonances. The observed modes originate from the hybridization of three plasmonic modes (in-plane dipolar, in-plane hexapolar and out-of-plane quadrupolar modes) of the elliptical nanocylinder and excitation of the lattice modes (in a substrate and superstrate) if these lattice resonances are supported by the plasmonic modes of the nanoparticles.

ZnO nanocrystals influence the position of plasmonic modes and increase the role of the superstrate medium which results in the appearance of two separate modes in the small spectral region. The enhancement of ZnO nanocrystals photoluminescence benefits from the simultaneous excitation and emission enhancements. Thereby we show that the analyte itself can play an important role in the plasmonic modes appearing due to its refractive index. Taking into account the possible benefits of the analyte refractive index may result in the optimization of the surface enhancing system. Therefore, the combination of the well-designed Al nanostructures and ZnO nanocrystals may play a key role in high-efficiency optoelectronic devices.

## Conflicts of interest

There are no conflicts to declare.

## Acknowledgements

The authors would like to thank the platform Nano'Mat (http://www. nanomat.eu), EIPHI Graduate School (contract "ANR17-EURE-0002") and acknowledge Eiffel scholarship and Belarussian Foundation for Fundamental Research. The numerical simulations were supported by the HPC Center of Champagne-Ardenne ROMEO.

## Notes and references

1   S. K. Jha, Z. Ahmed, M. Agio, Y. Ekinci and J. F. Löffler, *J. Am. Chem. Soc.*, 2012, **134**, 1966–1969.
2   G. V. Naik, V. M. Shalaev and A. Boltasseva, *Adv. Mater.*, 2013, **25**, 3264–3294.
3   Y. Ekinci, H. H. Solak and J. F. Löffler, *J. Appl. Phys.*, , DOI:10.1063/1.2999370.
4   M. W. Knight, N. S. King, L. Liu, H. O. Everitt, P. Nordlander and N. J. Halas, *ACS Nano*, 2013, **8**, 834–840.
5   D. Gerard and S. K. Gray, *J. Phys. D. Appl. Phys.*, 2015, **48**, 184001.
6   D. Khlopin, F. Laux, W. P. Wardley, J. Martin, G. A. Wurtz, J. Plain, N. Bonod, A. V. Zayats, W. Dickson and D. Gérard, *J. Opt. Soc. Am. B*, 2017, **34**, 691.
7   K. J. Smith, Y. Cheng, E. S. Arinze, N. E. Kim, A. E. Bragg and S. M. Thon, *ACS Photonics*, 2018, **5**, 805–813.
8   M. Schade, B. Fuhrmann, C. Bohley, S. Schlenker, N. Sardana, J. Schilling and H. S. Leipner, *J. Appl. Phys.*, 2014, **115**, 084309.
9   K. B. Mogensen, M. Gühlke, J. Kneipp, S. Kadkhodazadeh, J. B. Wagner, M. Espina Palanco, H. Kneipp and K. Kneipp, *Chem. Commun.*, 2014, **50**, 3744–3746.
10  M. Chowdhury, K. Ray, S. Gray, J. Pond and J. R. Lakowicz, *Anal. Chem.*, 2009, **81**, 1397–1403.
11  T. Dörfer, M. Schmitt and J. Popp, *J. Raman Spectrosc.*, 2007, **38**, 1379–1382.
12  D. O. Sigle, E. Perkins, J. J. Baumberg and S. Mahajan, *J. Phys. Chem. Lett.*, 2013, **4**, 1449–1452.
13  Y. Kawachiya, S. Murai, M. Saito, H. Sakamoto, K. Fujita and K. Tanaka, *Opt. Express*, 2018, **26**, 5970.
14  C. L. Lay, C. S. L. Koh, J. Wang, Y. H. Lee, R. Jiang, Y. Yang, Z. Yang, I. Y. Phang and X. Y. Ling, *Nanoscale*, 2018, **10**, 575–581.
15  S. Tian, O. Neumann, M. J. McClain, X. Yang, L. Zhou, C. Zhang, P. Nordlander and N. J. Halas, *Nano Lett.*, 2017, **17**, 5071–5077.
16  L. Cui, H. J. Butler, P. L. Martin-Hirsch and F. L. Martin, *Anal. Methods*, 2016, **8**, 481–487.
17  G. Lozano, D. J. Louwers, S. R. K. Rodríguez, S. Murai, O. T. A. Jansen, M. A. Verschuuren and J. Gómez Rivas, *Light Sci. Appl.*, 2013, **2**, e66.
18  V. G. Kravets, A. V. Kabashin, W. L. Barnes and A. N. Grigorenko,

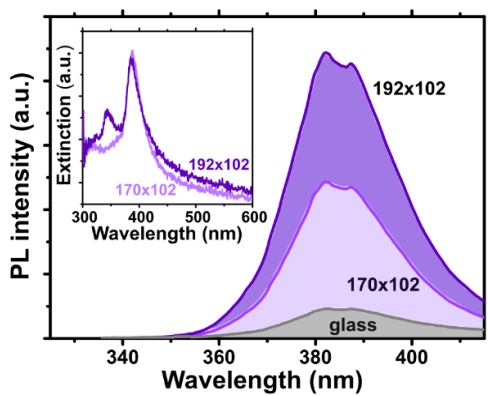

**Fig 8.** Photoluminescence spectra of a ZnO nanocrystals deposited on the array of 192 nm x 102 nm (dark violet) and 170 nm x 102 nm (light violet) Al nanocylinders and on a reference glass substrate (gray). Inset shows the extinction spectra of the two arrays.

*Chem. Rev.*, 2018, **118**, 5912–5951.
19  J. Marae-Djouda, R. Caputo, N. Mahi, G. Lévêque, A. Akjouj, P.-M. Adam and T. Maurer, *Nanophotonics*, 2017, **6**, 279–288.
20  M. Ross, C. Mirkin and G. Schatz, *J. Phys. Chem. C*, 2016, **120**, 816–830.
21  A. Yang, A. J. Hryn, M. R. Bourgeois, W.-K. Lee, J. Hu, G. C. Schatz and T. W. Odom, *Proc. Natl. Acad. Sci.*, 2016, **113**, 14201–14206.
22  P. P. Pompa, L. Martiradonna, A. Della Torre, F. Della Sala, L. Manna, M. de Vittorio, F. Calabi, R. Cinagolani and R. Rinaldi, *Nat. Nanotechnol.*, 2006, **1**, 126–130.
23  N. Mahi, G. Lévêque, O. Saison, J. Marae-Djouda, R. Caputo, A. Gontier, T. Maurer, P.-M. Adam, B. Bouhafs and A. Akjouj, *J. Phys. Chem. C*, 2017, **121**, 2388–2401.
24  Q. Y. Lin, Z. Li, K. A. Brown, M. N. O'Brien, M. B. Ross, Y. Zhou, S. Butun, P. C. Chen, G. C. Schatz, V. P. Dravid, K. Aydin and C. A. Mirkin, *Nano Lett.*, 2015, **15**, 4699–4703.
25  A. Muravitskaya, A. Movsesyan, S. Kostcheev and P.-M. Adam, *J. Opt. Soc. Am. B*, 2019, **36**, 65–70.
26  B. Luk'Yanchuk, N. I. Zheludev, S. A. Maier, N. J. Halas, P. Nordlander, H. Giessen and C. T. Chong, *Nat. Mater.*, 2010, **9**, 707–715.
27  E. Prodan, C. Radloff, N. J. Halas and P. Nordlander, *Science (80-. ).*, 2003, **302**, 419–422.
28  Y. Francescato, V. Giannini and S. A. Maier, *ACS Nano*, 2012, **6**, 1830–1838.
29  S. Zhang, K. Bao, N. J. Halas, H. Xu and P. Nordlander, *Nano Lett.*, 2011, **11**, 1657–1663.
30  P. Nordlander, C. Oubre, E. Prodan, K. Li and M. I. Stockman, *Nano Lett.*, 2004, **4**, 899–903.
31  H. Chen, L. Shao, T. Ming, K. C. Woo, Y. C. Man, J. Wang and H. Q. Lin, *ACS Nano*, 2011, **5**, 6754–6763.
32  P. Nordlander and E. Prodan, *Nano Lett.*, 2004, **4**, 2209–2213.
33  P. Spinelli, C. van Lare, E. Verhagen and A. Polman, *Opt. Express*, 2011, **19**, A303–A311.
34  C. Klingshirn, J. Fallert, H. Zhou, J. Sartor, C. Thiele, F. Maier-Flaig, D. Schneider and H. Kalt, *Phys. Status Solidi*, 2010, **247**, 1424–1447.
35  R. C. Monreal, S. P. Apell and T. J. Antosiewicz, *Nanoscale*, 2018, **10**, 7016–7025.
36  W. Z. Liu, H. Y. Xu, C. L. Wang, L. X. Zhang, C. Zhang, S. Y. Sun, J. G. Ma, X. T. Zhang, J. N. Wang and Y. C. Liu, *Nanoscale*, 2013, **5**, 8634–8639.
37  C. W. Lai, J. An and H. C. Ong, *Appl. Phys. Lett.*, 2005, **86**, 1–3.
38  H. Y. Lin, C. L. Cheng, Y. Y. Chou, L. L. Huang, Y. F. Chen and K. T. Tsen, *Opt. Express*, 2006, **14**, 2372–2379.
39  C. W. Cheng, E. J. Sie, B. Liu, C. H. A. Huan, T. C. Sum, H. D. Sun and H. J. Fan, *Appl. Phys. Lett.*, 2010, **96**, 071107.
40  W. H. Ni, J. An, C. W. Lai, H. C. Ong and J. B. Xu, *J. Appl. Phys.*, 2006, **100**, 026103.
41  K. Wu, Y. Lu, H. He, J. Huang, B. Zhao and Z. Ye, *J. Appl. Phys.*, 2011, **110**, 023510.
42  J. Lu, J. Li, C. Xu, Y. Li, J. Dai, Y. Wang, Y. Lin and S. Wang, *ACS Appl. Mater. Interfaces*, 2014, **6**, 18301–18305.
43  M. Norek, G. Łuka and M. Włodarski, *Appl. Surf. Sci.*, 2016, **384**, 18–26.
44  Y. Lin, X. Q. Liu, T. Wang, C. Chen, H. Wu, L. Liao and C. Liu, *Nanotechnology*, 2013, **24**, 125705.
45  J. Lu, C. Xu, J. Dai, J. Li, Y. Wang, Y. Lin and P. Li, *Nanoscale*, 2015, **7**, 3396–3403.
46  C. Stelling, C. R. Singh, M. Karg, T. König, M. Thelakkat and M. Retsch, *Sci. Rep.*, 2017, **7**, 1–13.
47  A. Movsesyan, A. L. Baudrion and P. M. Adam, *J. Phys. Chem. C*, 2018, **122**, 23651–23658.
48  P. Törmö and W. L. Barnes, *Reports Prog. Phys.*, 2015, **78**, 013901.
49  A.-L. Baudrion, A. Perron, A. Veltri, A. Bouhelier, P.-M. Adam and R. Bachelot, *Nano Lett.*, 2013, **13**, 282–286.
50  J. Liu, W. Wang, D. Wang, J. Hu, W. Ding, R. D. Schaller, G. C. Schatz and T. W. Odom, *Proc. Natl. Acad. Sci. U. S. A.*, 2019, **116**, 5925–5930.
51  J. Henson, J. Dimaria and R. Paiella, *J. Appl. Phys.*, 2009, **106**, 093111.
52  N. J. Halas, S. Lal, W. S. Chang, S. Link and P. Nordlander, *Chem. Rev.*, 2011, **111**, 3913–3961.
53  E. J. Smythe, E. Cubukcu and F. Capasso, *Opt. Express*, 2007, **15**, 7439.
54  A. G. Nikitin, T. Nguyen and H. Dallaporta, *Appl. Phys. Lett.*, 2013, **102**, 221116.
55  N. Félidj, G. Laurent, J. Aubard, G. Lévi, A. Hohenau, J. R. Krenn and F. R. Aussenegg, *J. Chem. Phys.*, 2005, **123**, 221103.
56  T. R. Jensen, M. L. Duval, K. L. Kelly, A. A. Lazarides, G. C. Schatz and R. P. Van Duyne, *J. Phys. Chem. B*, 1999, **103**, 9846–9853.
57  K. M. Mayer and J. H. Hafner, *Chem. Rev.*, 2011, **111**, 3828–3857.
58  I. Kaminska, T. Maurer, R. Nicolas, M. Renault, T. Lerond, R. Salas-Montiel, Z. Herro, M. Kazan, J. Niedziolka-Jönsson, J. Plain, P.-M. Adam, R. Boukherroub and S. Szunerits, *J. Phys. Chem. C*, 2015, **119**, 9470–9476.
59  A. F. Koenderink, *ACS Photonics*, 2017, **4**, 710–722.
60  D. V. Guzatov, S. V. Vaschenko, V. V. Stankevich, A. Y. Lunevich, Y. F. Glukhov and S. V. Gaponenko, *J. Phys. Chem. C*, 2012, **116**, 10723–10733.
61  S. V. Gaponenko, P.-M. Adam, D. V. Guzatov and A. Muravitskaya, *Sci. Rep.*, 2019, **9**, 7138.